\documentclass{article}
\usepackage{spconf,amsmath,graphicx,hyperref}
\usepackage{float}
\title{DPC-QA Net: A No-Reference Dual-Stream Perceptual and Cellular Quality Assessment Network for Histopathology Images}
\name{Qijun Yang$^{1}$, Boyang Wang$^{2}$,  Hujun Yin$^{1}$}
\address{$^{1}$ DoEEE, University of Manchester  \quad $^{2}$  CCSE, Beihang University }

\begin{document}
%
\maketitle

\begin{abstract}
Reliable whole slide imaging (WSI) hinges on image quality, yet staining artefacts, defocus, and cellular degradations are common. We present DPC\mbox{-}QA Net, a no\mbox{-}reference dual\mbox{-}stream network that couples wavelet\mbox{-}based global difference perception with cellular quality assessment from nuclear and membrane embeddings via an Aggr\mbox{-}RWKV module. Cross\mbox{-}attention fusion and multi\mbox{-}term losses align perceptual and cellular cues. Across different datasets, our model detects staining, membrane, and nuclear issues with $>$92\% accuracy and aligns well with usability scores; on LIVEC and KonIQ it outperforms state\mbox{-}of\mbox{-}the\mbox{-}art NR\mbox{-}IQA. A downstream study further shows strong positive correlations between predicted quality and cell recognition accuracy (e.g., nuclei PQ/Dice, membrane boundary F\mbox{-}score), enabling practical pre\mbox{-}screening of WSI regions for computational pathology.
\end{abstract}


\section{Introduction}

Whole slide imaging (WSI) is central to digital pathology, enabling large-scale studies, telepathology, and AI-assisted diagnosis. Its reliability, however, depends critically on image quality, which may be degraded by staining artefacts, scanner defocus, and cellular distortions such as blurred nuclei or membranes. Low-quality slides can significantly hinder computational pathology and diagnostic reproducibility~\cite{janowczyk2019histoqc}. Unlike natural images, WSIs contain highly repetitive structures and complex distortions, making conventional no-reference image quality assessment (NR-IQA) methods unsuitable~\cite{guo2023blind,hosseini2020focuspath}. Manual inspection and rule-based measures remain widely used but are subjective and inefficient~\cite{guo2020focuslite}.

Medical IQA research highlights modality-specific needs, with dedicated solutions for MRI, CT, ultrasound, X-ray, and PET~\cite{hu2022chest,hopson2020fast}. Pathology introduces unique challenges, where tissue staining and preparation cause artefacts beyond other modalities. Early focus quality metrics (e.g.,  FQPath~\cite{hosseini2020fqpath}) and open-source tools like CellProfiler~\cite{mcquin2018cellprofiler} and HistoQC~\cite{janowczyk2019histoqc} paved the way, while deep learning frameworks~\cite{martin2022automated,guo2020focuslite,zhou2025perceptual} demonstrated scalability to large cohorts.  

Meanwhile, NR-IQA has advanced with natural-scene models (BRISQUE~\cite{mittal2012brisque}),  and recent deep approaches including quality-difference learning~\cite{wu2022qualitydiff}, panoramic IQA~\cite{pan2021panoramic} and Transformers~\cite{you2021transformer}. Pathology-specific adaptations~\cite{guo2023blind} introduced saliency-guided and patch-interaction networks. Yet most approaches treat either global or local cues, failing to address the \textbf{macro--micro inconsistency} in WSIs, where global staining appears acceptable but cellular detail is degraded.

To address this, we propose \textbf{DPC-QA Net}, a dual-stream NR-IQA framework for histopathology. It consists of:  
1) a \textbf{global perceptual stream} using wavelet-based difference-aware features to capture staining consistency and tissue completeness, and  
2) a \textbf{cellular quality stream} leveraging segmentation-derived embeddings to assess nuclear and membrane fidelity.  

The two streams are fused via gated cross-attention and discrepancy-aware regression, producing a comprehensive score aligned with both global and cellular fidelity. DPC-QA Net also generates interpretable heatmaps localising staining and cellular artefacts. Contributions as follow:
\begin{itemize}
    \item We propose the first dual-stream NR-IQA framework to address \textbf{macro--micro perceptual inconsistency} in WSIs.  
    \item The global stream employs \textbf{wavelet-based difference-aware features}, while the cellular stream introduces an \textbf{Aggr-RWKV module} to aggregate nuclear and membrane embeddings for fine-grained cell structure perception.   
\end{itemize}

\section{Proposed Method}

\begin{figure*}[t]
    \centering
    \includegraphics[width=0.8\textwidth]{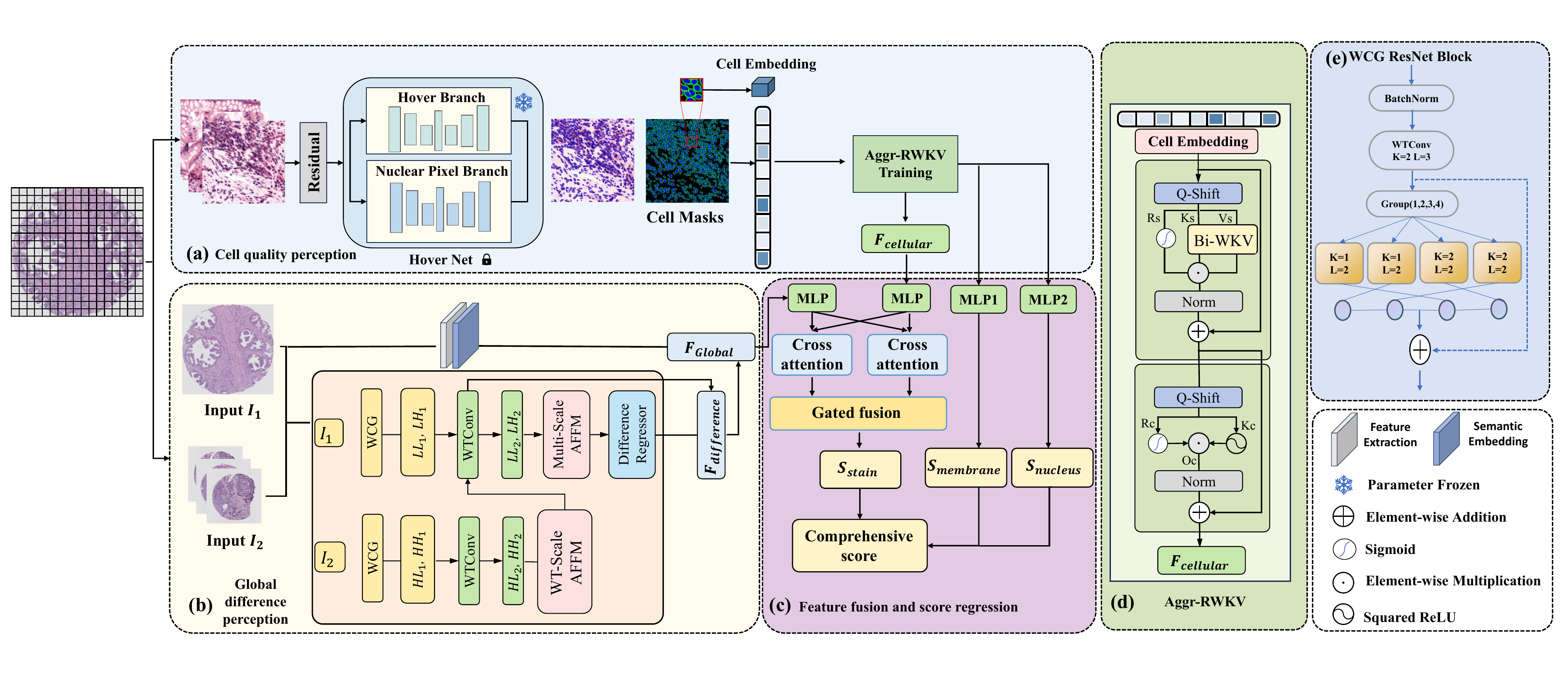}
    \caption{Pipeline of DPC-QA Net. A dual-stream design integrates (i) global difference perception with WCG-based feature extraction and (ii) cellular quality perception via Hover-Net embeddings. Cross-attention and gated fusion yield a final regressed quality score.}
    \label{fig:pipeline}
\end{figure*}

We propose \textbf{DPC-QA Net}, a dual-stream difference-perception and cellular-quality aware network for no-reference pathological image quality assessment. As illustrated in Figure.~\ref{fig:pipeline}, the framework consists of a global difference perception branch, a cellular quality perception branch, and a cross-branch fusion and regression module that integrates both feature streams into a comprehensive staining quality score.  

\textbf{Global Difference Perception Branch.}  
Given an input patch $I$, we perform wavelet decomposition to obtain low- and high-frequency sub-bands:  
\begin{equation}
[X_{LL}, X_{LH}, X_{HL}, X_{HH}] = \text{WT}(I).
\end{equation}
The sub-bands are processed by Wavelet-Convolutional Group (WCG) blocks, followed by inverse wavelet reconstruction:
\begin{equation}
Y = \text{IWT}\left(\text{DWConv}(X_{LL}), \{\text{DWConv}(X_h)\}_{h \in \{LH,HL,HH\}}\right).
\end{equation}
Multi-scale asymmetric feature fusion module (AFFM) aggregates these representations:
\begin{equation}
\mathbf{F}_{difference} = \text{AFFM}\big(\{Y_1, Y_2\}\big).
\end{equation}
To capture long-range dependencies, we employ a Bi-WKV module with Q-shift normalization:
\begin{equation}
\mathbf{F}_{Global} = \text{Bi-WKV}(\mathbf{F}_{difference}).
\end{equation}

\textbf{Cellular Quality Perception Branch.}  
Hover-Net provides nuclear masks $M_{\text{nuc}}$. We then generate membrane masks $M_{\text{mem}}$ by dilating $M_{\text{nuc}}$ and subtracting its interior. Based on these masks, we extract:  
\begin{equation}
\mathbf{S}_{nucleus} = \phi_{\text{nuc}}(I \odot M_{\text{nuc}}), \quad
\mathbf{S}_{membrane} = \phi_{\text{mem}}(I \odot M_{\text{mem}}),
\end{equation}
where $\phi_{\text{nuc}}(\cdot)$ and $\phi_{\text{mem}}(\cdot)$ are lightweight CNN encoders. The nuclear score $\mathbf{S}_{nucleus}$ evaluates chromatin clarity, while the membrane score $\mathbf{S}_{membrane}$ reflects cell boundary integrity.  

These features are aggregated using \textbf{Aggregation via Receptance-Weighted Key-Value (Aggr-RWKV)}. Unlike conventional self-attention ($O(n^2)$), Aggr-RWKV achieves linear complexity $O(n)$, making it scalable to thousands of cells per patch. Formally,  
\begin{equation}
\mathbf{F}_{cellular} = \text{Aggr-RWKV}(\mathbf{S}_{membrane} \Vert \mathbf{S}_{nucleus}).
\end{equation}

\textbf{Cross-Branch Fusion and Regression.}  
We fuse $\mathbf{F}_{cellular}$ and $\mathbf{F}_{Global}$ via cross-attention:
\begin{equation}
F_{\text{fusion}} = \text{softmax}\left(\frac{QK^\top}{\sqrt{d}}\right)V,
\end{equation}
where $Q=\mathbf{F}_{cellular}, K=V=\mathbf{F}_{Global}$. Gated fusion balances contributions:  
\begin{equation}
F_{\text{fused}} = \sigma(W[F_{\text{fusion}} \Vert \mathbf{F}_{cellular}]) \odot F_{\text{fusion}} + (1-\sigma(\cdot)) \odot \mathbf{F}_{cellular}.
\end{equation}
Finally, the MLP regressor outputs the staining quality score:  
\begin{equation}
\mathbf{S}_{stain} = \text{MLP}(F_{\text{fused}}).
\end{equation}

\textbf{Loss Function.}  
The network is trained with:  
\begin{equation}
\mathcal{L} = \mathcal{L}_{\text{reg}} + \lambda_1 \mathcal{L}_{\text{diff}} + \lambda_2 \mathcal{L}_{\text{wavelet}} + \lambda_3 \mathcal{L}_{\text{Aggr}},
\end{equation}
where $\mathcal{L}_{\text{reg}} = \|\mathbf{S}_{stain} - S^\ast\|_1$,  
$\mathcal{L}_{\text{diff}} = \|d - |\mathbf{S}_{stain}^i - \mathbf{S}_{stain}^j|\|_1$,  
$\mathcal{L}_{\text{wavelet}} = \sum_{j,k}\|\mathcal{W}_{j,k}(I) - \mathcal{W}_{j,k}(\hat{I})\|_1$,  
and $\mathcal{L}_{\text{Aggr}}$ regularizes aggregation consistency.  

The slide-level score is obtained by averaging patch predictions:  
\begin{equation}
S_{\text{WSI}} = \frac{1}{N}\sum_{i=1}^N \mathbf{S}_{stain}^i.
\end{equation}

\section{Experiments}
\subsection{Experimental Setup}

\begin{figure*}[t]
    \centering
    \includegraphics[width=0.8\textwidth]{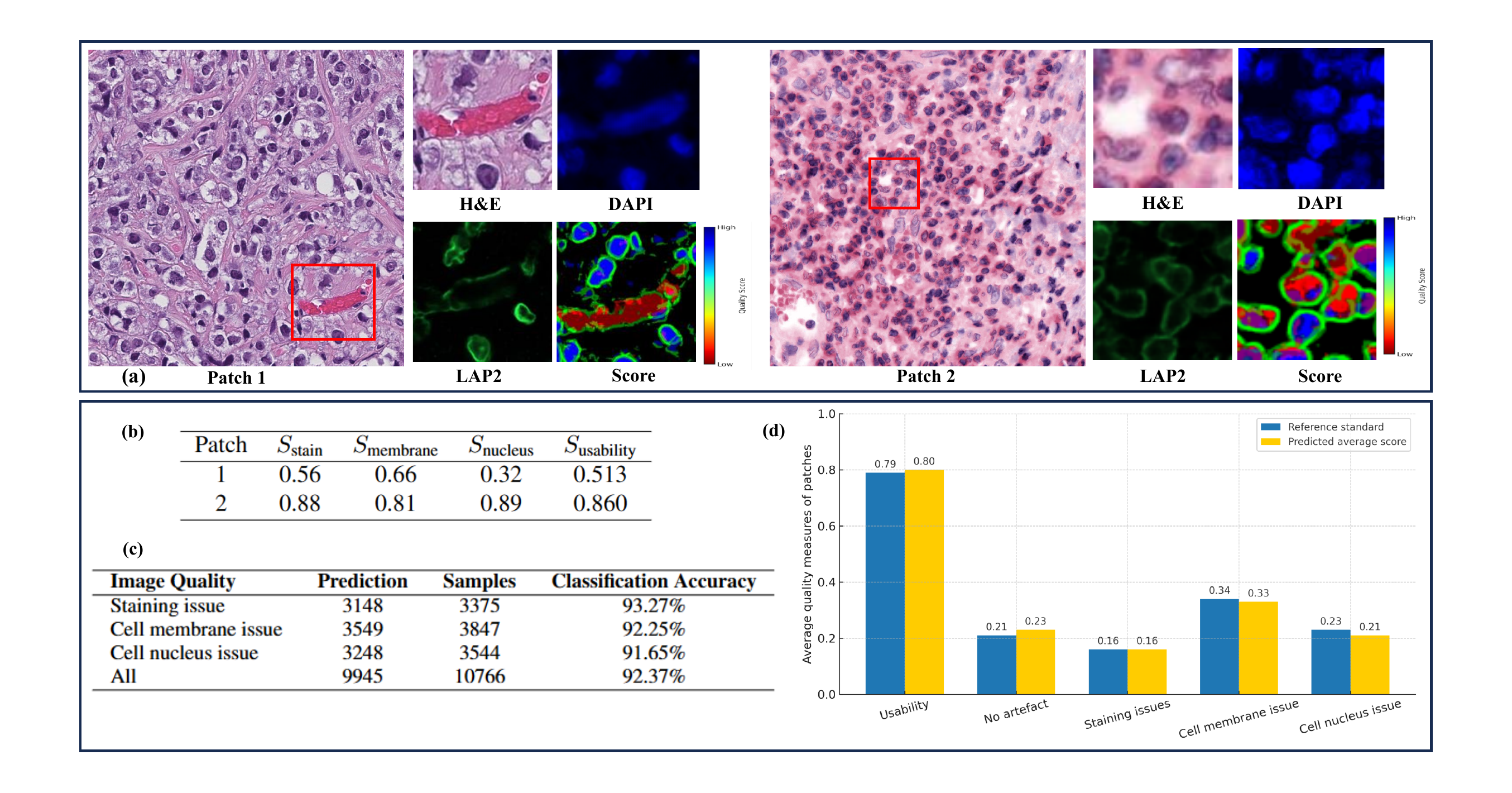}
    \caption{(a) Representative patches (Patch 1 and Patch 2) extracted from whole-slide images (WSIs) of rectal cancer tissue. (b) Evaluation scores of the two patches across staining, cell membrane, cell nucleus, and usability metrics. (c) Prediction accuracy of HEMIT for staining, cell membrane, and cell nucleus quality issues. (d) Comparison between predicted and reference average quality scores on the HEMIT dataset.}
    \label{fig:Illustration of the image quality assessment results}
\end{figure*}
\vspace{-2pt}

A retrospective cohort of colorectal cancer WSIs was digitised at high resolution and divided into non-overlapping patches of size $256 \times 256$ at 5$\times$ magnification, yielding 10,766 patches. All samples were annotated by board-certified colorectal pathologists on the HEMIT platform~\cite{bian2024hemit}, reporting one or more of the following quality issues: \textit{staining artefacts} (3,375), \textit{membrane blur} (3,847), and \textit{nuclear degradation} (3,544), with overlaps allowed. To enhance visibility of subcellular structures, virtual LAP2 and DAPI staining were generated under expert guidance. Each patch was also assigned a binary usability label and a continuous score in $[0,1]$. The dataset was split into 70\% training (7,536), 10\% validation (1,075), and 20\% testing (2,155), ensuring no patch overlap across sets.  
\vspace{-1pt}
All experiments were conducted in PyTorch 1.12 on an NVIDIA RTX 4090 GPU (24 GB). Models were trained with the Adam optimizer (default parameters), an initial learning rate of $1\times10^{-4}$, weight decay $1\times10^{-5}$, batch size 16, and up to 100 epochs with early stopping. Hidden feature dimensions were fixed at 256. The overall training loss was a weighted sum of regression, difference-perception, wavelet, and aggregation terms, as detailed in Table~\ref{1}.  

\vspace{-6pt}
\begin{table}[t]
\centering
\caption{Loss components and corresponding weights used in training.}
\scriptsize
\resizebox{\columnwidth}{!}{
\begin{tabular}{lcc}
\hline
\textbf{Loss Term} & \textbf{Definition} & \textbf{Weight} \\
\hline
$L_{\text{reg}}$ & $\| S - S^{*} \|_{1}$ & 1.0 \\
$L_{\text{diff}}$ & $\| d - |S_i - S_j| \|_{1}$ & $\lambda_{1}=0.5$ \\
$L_{\text{wavelet}}$ & $\sum_{j,k} \| W_{j,k}(I) - W_{j,k}(\hat{I}) \|_{1}$ & $\lambda_{2}=0.1$ \\
$L_{\text{Aggr}}$ & Aggregation consistency & $\lambda_{3}=0.5$ \\
\hline
\end{tabular}}
\label{1}
\end{table}
\vspace{-6pt}

\subsection{Results}

On the HEMIT dataset, our method achieved accuracies of 93.27\%, 92.25\%, and 91.65\% in detecting staining, membrane, and nuclear artefacts, respectively, with predicted scores closely matching expert annotations (Figure.~\ref{fig:Illustration of the image quality assessment results}). This demonstrates strong robustness across different artefact types.  

We also conducted cross-dataset evaluation using ProMPT \cite{martin2022automated} (prostate cancer, FFPE), TCGA~\cite{cooper2018pancancer} (prostate cancer, mainly frozen sections), and HEMIT~\cite{bian2024hemit} (colorectal cancer). Table~\ref{tab:patch_quality} shows that DPC-QA Net maintains high agreement with reference scores, achieving near-perfect usability prediction ($\geq$0.98) and stable artefact detection accuracies ($>0.81$) even on external cohorts. These results highlight the robustness and transferability of the proposed framework across diverse WSI datasets.  

To further assess domain-transfer capability, we benchmarked on the natural-image NR-IQA datasets LIVEC and KonIQ.
As shown in Table~\ref{tab:comparison}, classical NSS-based models (e.g., BRISQUE, BLIINDS-II) underperform compared with recent deep methods on these benchmarks.
Despite being tailored for histopathology, our framework attains the highest PLCC/SRCC on both datasets, indicating that its perceptual modelling generalises beyond WSI content and does not overfit to tissue-specific statistics.

\begin{table}[t]
\centering
\caption{Average patch-level quality measurements across different WSI datasets (ProMPT \cite{martin2022automated}, TCGA \cite{cooper2018pancancer}, HEMIT \cite{bian2024hemit}).}
\footnotesize
\setlength{\tabcolsep}{3pt}
\begin{tabular}{lcccccc}
\hline
Metric & \multicolumn{2}{c}{ProMPT} & \multicolumn{2}{c}{TCGA} & \multicolumn{2}{c}{HEMIT} \\
       & Pred./Ref. & Acc. & Pred./Ref. & Acc. & Pred./Ref. & Acc. \\
\hline
Usability       & 0.889/0.869 & 0.980 & 0.733/0.751 & 0.982 & 0.802/0.798 & 0.996 \\
Staining issues & 0.163/0.165 & 0.998 & 0.085/0.102 & 0.983 & 0.165/0.166 & 0.999 \\
Cell membrane   & 0.433/0.622 & 0.811 & 0.382/0.451 & 0.931 & 0.334/0.341 & 0.993 \\
Cell nucleus    & 0.421/0.579 & 0.842 & 0.352/0.469 & 0.883 & 0.215/0.226 & 0.989 \\
No artefact     & 0.233/0.192 & 0.959 & 0.114/0.149 & 0.965 & 0.361/0.365 & 0.996 \\
\hline
\end{tabular}
\label{tab:patch_quality}
\end{table}

\begin{table}[htbp]
\centering
\caption{Performance comparison of results across two real distortion datasets}
\footnotesize
\setlength{\tabcolsep}{4pt} 
\begin{tabular}{l|cc|cc}
\hline
\textbf{METHOD} & \multicolumn{2}{c}{\textbf{LIVEC}} & \multicolumn{2}{c}{\textbf{KonIQ}} \\
\cline{2-3} \cline{4-5}
& PLCC & SRCC & PLCC & SRCC \\
\hline
BLIINDS-II \cite{saad2012blind} & 0.4077 & 0.2267 & 0.3848 & 0.1534 \\
BRISQUE \cite{zhang2015feature}   & 0.6088 & 0.6094 & 0.2556 & 0.1426 \\
\hline
MetaIQA \cite{zhu2020metaiqa}    & 0.8069 & 0.8367 & 0.8628 & 0.8757 \\
HyperIQA \cite{su2020blindly}  & 0.8919 & 0.8631 & 0.9078 & 0.9162 \\
IQDLNet \cite{xie2023no} & 0.8751 & 0.8685 & 0.9212 & 0.9059 \\
ProMPT \cite{haghighat2022automated}      & 0.8784 & 0.8725 & 0.9049 & 0.9175 \\
\hline
\textbf{Proposed}   & \textbf{0.8933} & \textbf{0.8774} & \textbf{0.9277} & \textbf{0.9192} \\
\hline
\end{tabular}
\label{tab:comparison}
\end{table}

\vspace{-5pt}

\subsection{Downstream analysis: do higher quality scores yield better cell recognition?}

\begin{figure}[t]
    \centering
    \includegraphics[width=\columnwidth]{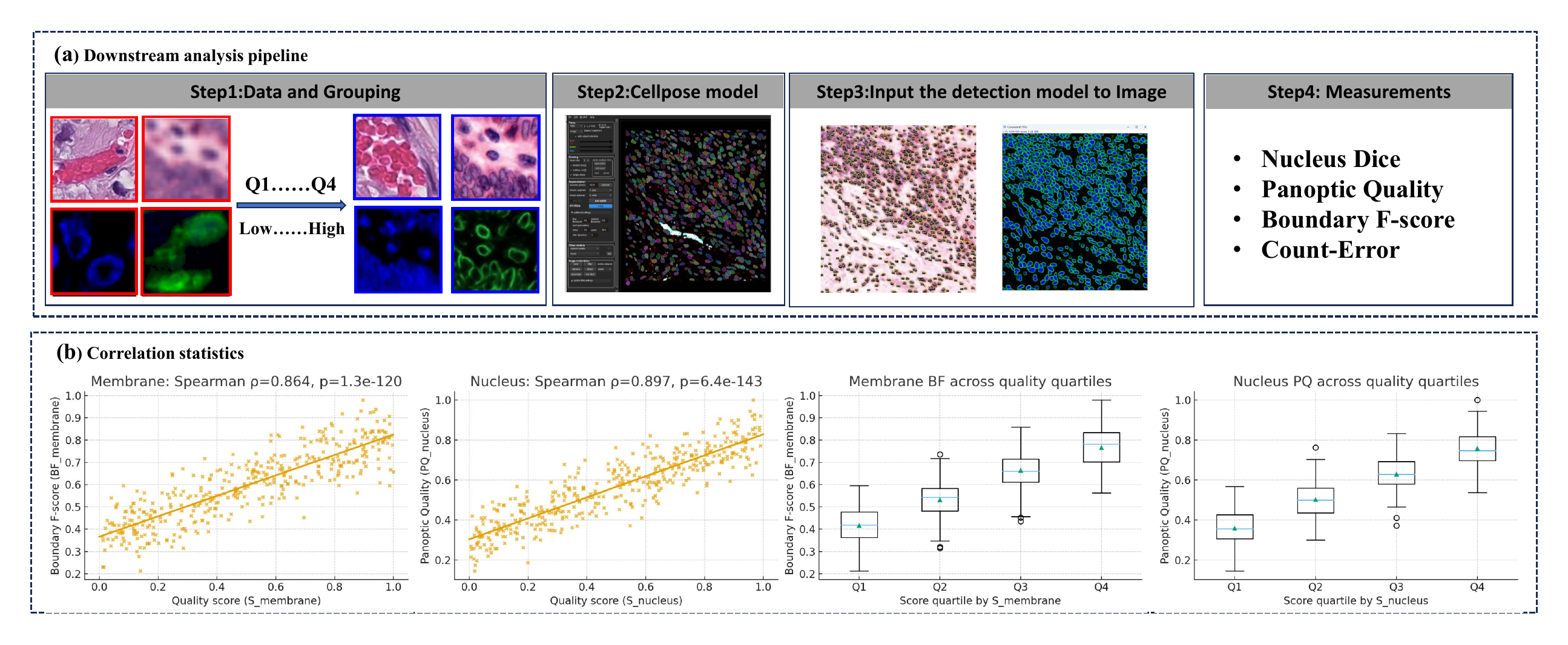}
    \caption{Downstream validation of DPC\mbox{-}QA for cell recognition. 
    \textbf{(a)} Workflow: score patches $\rightarrow$ fixed Cellpose inference $\rightarrow$ compute metrics. 
    \textbf{(b)} Higher scores associate with better nuclei PQ and membrane BF; quartiles (Q1$\rightarrow$Q4) show monotonic gains. 
    Statistics: Spearman $\rho$ and Kruskal--Wallis; see Table~\ref{tab:quality_corr}.}
    \label{fig:downstream_quality}
\end{figure}

\begin{table}[t] 
\centering
\caption{Correlation between DPC\mbox{-}QA scores and cell recognition accuracy on HEMIT.
$\rho$: Spearman correlation; $p$: two-sided $p$-value.
G1--G4: median metric values by $S_{\text{usability}}$ score bins (G1: $[0,0.2)$, G2: $[0.2,0.4)$, G3: $[0.4,0.6)$, G4: $[0.6,1.0]$).}
\footnotesize   
\setlength{\tabcolsep}{3pt} 
\begin{tabular}{lcccccc}
\hline
\textbf{Metric} & $\rho$ & $p$ & G1 & G2 & G3 & G4 \\
\hline
Nucleus Dice      & 0.936 & $6.3\times10^{-183}$ & 0.398 & 0.551 & 0.673 & 0.816 \\
Panoptic Quality  & 0.897 & $6.4\times10^{-143}$ & 0.357 & 0.500 & 0.629 & 0.747 \\
Boundary F-score  & 0.864 & $1.3\times10^{-120}$ & 0.419 & 0.544 & 0.660 & 0.782 \\
Count-Error (↓)   & -0.743 & $2.1\times10^{-71}$  & 0.217 & 0.182 & 0.137 & 0.087 \\
\hline
\end{tabular}
\label{tab:quality_corr}
\end{table}

We assess whether DPC\mbox{-}QA scores predict downstream cell recognition reliability. 
From the HEMIT annotated set we randomly sampled patches from several WSIs, forming \textbf{sets of 300 patches} each (no slide overlap across sets). 
Patches are then grouped by the DPC\mbox{-}QA \textbf{usability score} $S_{\text{usability}}$ into four bins to avoid slide\mbox{-}level bias: 
\textbf{G1} $[0,0.2)$, \textbf{G2} $[0.2,0.4)$, \textbf{G3} $[0.4,0.6)$, \textbf{G4} $[0.6,1.0]$. 
A \emph{fixed} pretrained Cellpose pipeline (no fine\mbox{-}tuning) is applied, and we compute instance\mbox{-}/pixel\mbox{-}level metrics (nuclei PQ/Dice, membrane boundary F\mbox{-}score) and count\mbox{-}error.

\textbf{Findings.} 
As summarised in Table~\ref{tab:quality_corr} and illustrated in Fig.~\ref{fig:downstream_quality}, quality scores show strong positive correlations with recognition accuracy (e.g., $S_{\text{nuc}}$ vs.\ nuclei PQ/Dice; $S_{\text{mem}}$ vs.\ membrane BF) and a negative correlation with count\mbox{-}error. 
Groupwise analysis reveals a clear monotonic improvement from \textbf{G1} (lowest\mbox{-}quality) to \textbf{G4} (highest\mbox{-}quality), confirmed by non\mbox{-}parametric tests. 
These results indicate that the proposed scores—especially the cellular stream outputs aggregated by Aggr\mbox{-}RWKV—are informative proxies for downstream performance and can pre\mbox{-}screen WSI regions to reduce segmentation failures in computational pathology workflows.

\subsection{Ablation Study}
\vspace{-6pt}
\begin{table}[H]
\centering
\caption{Ablation study of DPC-QA Net on the HEMIT dataset.}
\resizebox{\columnwidth}{!}{
\begin{tabular}{lcc}
\hline
\textbf{Variant} & \textbf{PLCC} & \textbf{SRCC} \\
\hline
Full DPC-QA Net (ours) & \textbf{0.9277} & \textbf{0.9192} \\
w/o WCG (replace with CNN conv) & 0.9012 & 0.8895 \\
w/o Aggr-RWKV (replace with avg pooling) & 0.8958 & 0.8836 \\
w/o Cross-attention fusion & 0.8874 & 0.8741 \\
w/o $L_{\text{diff}}$ & 0.9086 & 0.8927 \\
w/o $L_{\text{wavelet}}$ & 0.9125 & 0.8971 \\
w/o $L_{\text{Aggr}}$ & 0.9153 & 0.9012 \\
\hline
\end{tabular}}
\label{tab:ablation}
\end{table}
\vspace{-6pt}

We further analysed the contribution of individual components on HEMIT (Table~\ref{tab:ablation}). Removing the \textbf{WCG blocks} in the global stream or the \textbf{Aggr-RWKV module} in the cellular stream leads to clear correlation drops, showing their effectiveness in modelling global artefacts and fine-grained nuclear features. Excluding the \textbf{cross-attention fusion} results in the largest degradation, underlining the importance of jointly modelling global and cellular cues. Eliminating individual loss terms ($L_{\text{diff}}$, $L_{\text{wavelet}}$, $L_{\text{Aggr}}$) causes moderate reductions, indicating complementary supervision. Overall, the ablation confirms the necessity of both architectural modules and multi-term loss design.

\section{Conclusion}
We proposed DPC-QA Net, a dual-stream NR-IQA framework that integrates wavelet-based global difference perception with cellular embeddings to address macro--micro quality inconsistencies in WSIs. Experiments on HEMIT, external cohorts, and natural-image benchmarks demonstrate high accuracy, robustness, and transferability, while ablation studies validate the contribution of each component. The framework provides an efficient and interpretable solution for automated WSI quality control, supporting reproducible diagnosis and scalable computational pathology workflows.

\vfill\pagebreak

\bibliography{refs}

\bibliographystyle{IEEEbib}

\end{document}